\documentclass[prl,twocolumn,amsmath,amssymb,groupedaddress,superscriptaddress]{revtex4-1}
\usepackage{graphics,color}
\usepackage{amsmath}
\usepackage[dvips]{graphicx}
\usepackage{float,graphicx}
\usepackage{epstopdf}
\newcommand{\beq}{\begin{equation}}
\newcommand{\eeq}{\end{equation}}
\newcommand{\beqa}{\begin{eqnarray}}

\newcommand{\eeqa}{\end{eqnarray}}
\def\be{\begin{equation}}

\def\ee{\end{equation}}

\def\iint{\textmd{int}}

\def\ttot{\textmd{tot}}
\def\tth{\textmd{th}}

\def\aaux{\textmd{aux}}

\def\Rre{\textmd{Re}}

\def\Ttr{\textmd{Tr}}

\def\be{\begin{equation}}
\def\ee{\end{equation}}
\def\bea{\begin{eqnarray}}
\def\eea{\end{eqnarray}}

\usepackage{bm}
\usepackage{graphicx}

\begin{document}
\bibliographystyle{unsrt,prl}
\title[Thermofield]{Thermofield-based chain mapping approach for open quantum systems}
\author{In{\'e}s de Vega}
\affiliation{Department of Physics and Arnold Sommerfeld Center for Theoretical Physics, Ludwig-Maximilians-University Munich, Germany}
\author{Mari-Carmen Ba\~nuls}
\affiliation{Max-Planck-Institut f\"ur Quantenoptik, Hans-Kopfermann-Str. 1, 85748 Garching, Germany.}

\begin{abstract}
We consider a thermofield approach to analyze the evolution of an open quantum system coupled to an environment at finite temperature. In this approach, the finite temperature environment is exactly mapped onto two virtual environments at zero temperature. These two environments are then unitarily transformed into two different chains of oscillators, leading to a one dimensional structure that can be numerically studied 
using tensor network techniques. 
\end{abstract}
\maketitle
In the past decades, many different techniques have been developed to analyze the dynamics of quantum systems coupled to an environment, i.e. open quantum systems (OQS). Some of these are based on deriving a master equation, 
which evolves the reduced density operator of the OQS by tracing out the environment degrees of freedom \cite{breuerbook,rivas2011a}, and some others are based on the stochastic Schr{\"o}dinger equations (SSE), evolving the OQS's wave function conditioned by a continuous \cite{diosi1998,alonso2005} or discrete \cite{plenio1998,piilo2009} stochastic process.
Both approaches are suitable for weak system-environment couplings, which generally lead to a large separation between system and environment time scales. 
Although such a large separation often occurs in quantum optics, it is not necessarily so in other scenarios, such as soft or condensed matter systems, or in quantum biology.
In these situations, other approaches are more appropriate, such as
the path integral Montecarlo \cite{wallsbook}, which in some parameter regimes is nevertheless hindered by the sign problem,
potentially affecting the convergence of the method at relatively short times (see for instance \cite{muelbacher2005}).

An alternative is to solve the total system dynamics with exact diagonalization methods,
which is difficult due to the large number of degrees of freedom in the environment.
 Hence, a wise selection of the relevant states of the full system is of primary importance, and this can be done for instance by discarding states with low probability, as in the density matrix approach \cite{zhang1998b} (closely related to density matrix renormalization group), or by considering as relevant only those states generated during the evolution, as done in the variational approach \cite{bonca1999,vidmar2010}.

Similarly, it is possible to perform a unitary transformation of the environment that \textit{maps} it onto a one dimensional structure. 
The numerical renormalization group (NRG) approach \cite{wilson1975,vojta2005,bulla2008,anders2007,weichselbaum2007,hughes2009}, for instance, 
is based on a (logarithmic) coarse-graining of the continuous environment spectral function in energy space. The resulting discretized environment can then be mapped onto a semi-infinite tight-binding chain \cite{krishna1980} with exponentially decreasing couplings. 
As proposed in \cite{prior2010,chin2010b,chinbook2011}, the mapping 
can also be performed analytically without a previous discretization of the environment.
Even when the couplings do not decay exponentially, it is typically possible to describe the system dynamics until its decay or relaxation time using a truncated chain of finite length.
The total system can now be modelled as a matrix product state (MPS), and it is then possible to use tensor network techniques to simulate
the unitary evolution of the total system \cite{vidal2003,verstraete08algo,scholl2011,scholl2005}.
The approach can also deal with an environment at finite temperature, using matrix product operators \cite{verstraete2004,zwolak2004}.

\begin{figure}[ht]
\centerline{\includegraphics[width=0.4\textwidth]{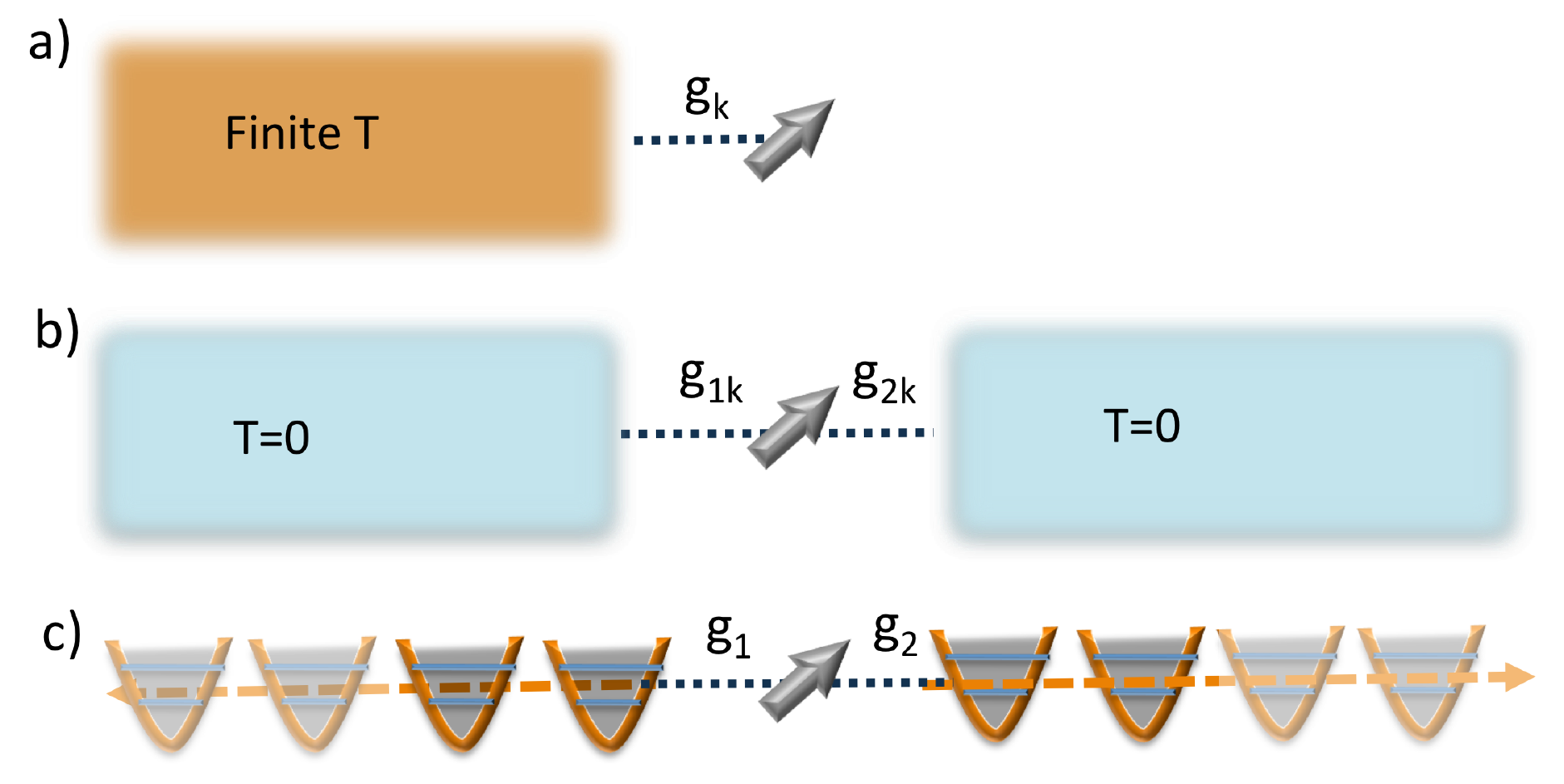}}
\caption{Fig. a) represents the initial problem described with (\ref{h1}) of an OQS coupled to a harmonic oscillator reservoir at finite temperature. Fig. b) represents the thermofield-transformed problem (\ref{h2}), in which the finite temperature of the reservoir is encoded in two different reservoirs at zero temperature. Fig. c) is the chain representation of the latter.} 
\label{thermofig}
\end{figure}

In this letter we present  a complementary formulation of this idea based on the thermofield approach proposed in \cite{bargmann1961,araki1963,takahashi1975} (see \cite{blasonebook} for a review). In this approach, the environmental Hilbert space is mirrored or doubled, and then a thermal Bogoliubov transformation is performed. As a result, the real environment in a thermal state is transformed into two virtual environments in a vacuum state (see Fig. \ref{thermofig}.b), known in the literature as the thermofield vacuum. The expectation value of any operator of the real environment in the thermofield vacuum coincides with its expectation value in the thermal state. The only excitations appearing in the environment will be those that are dynamically created through the interaction, and the dynamics of the resulting transformed system can 
 be simulated using MPS.

The thermofield approach has been considered in the framework of SSE of OQS (see for instance \cite{diosi1998}), but most of its applications are in the context of quantum field theory and general relativity \cite{maldacena2013,israel1976}. 

\textit{Thermofield dynamics--}
Let us consider an environment of harmonic oscillators, with annihilation (creation) operators $b_k$ ($b_k^{\dagger}$) and frequencies $\omega_k$,
to which the OQS 
couples with strengths $g_k$.
The complete Hamiltonian can be written as
\bea
H_{\ttot}=H_S +\sum_k \omega_{k} b^\dagger_k b_k +\sum_k g_k (L^{\dagger}b_k +b_k^{\dagger}L ),
\label{h1}
\eea
where $H_S$ is the Hamiltonian of the OQS, and $L$ is the coupling operator acting on the OQS Hilbert space. 
We can introduce an auxiliary, decoupled environment, 
 characterized by annihilation (creation) operators $c_k$ ($c_k^\dagger$) and write the total Hamiltonian as
\begin{eqnarray}
\hat{H}_{\ttot}&=&H_{\ttot}
-\sum_k \omega_{k} c^\dagger_k c_k.
\label{h2}
\end{eqnarray}
Assuming now that both environments 
are initially in a thermal state at inverse temperature $\beta$, we apply 
a thermal Bogoliubov transformation,
\bea
a_{1k}&=&e^{-iG}b_ke^{iG}=\cosh(\theta_{k})b_k-\sinh(\theta_{k})c^\dagger_k,\cr
a_{2k}&=&e^{-iG}c_ke^{iG}=\cosh(\theta_{k})c_k-\sinh(\theta_{k}) b^\dagger_k.
\label{bogoliub}
\eea
Here, $G=i\sum_k\theta_{k}(b_k^\dagger c^\dagger_k-c_kb_k)$, with $\theta_{k}$ a function of the temperature such that $\cosh(\theta_{k})=\sqrt{1+n_k}$, and $n_k=1/(e^{\beta\omega_{k}}-1)$ 
is the number of excitations in  mode $k$. 
In terms of these new modes,
\begin{eqnarray}
&&\hat{H}_{\ttot}=H_S +\sum_k \omega_{k} \large(a_{1k}^\dagger a_{1k}-a^\dagger_{2k}a_{2k}\large)\cr
&+&\sum_kg_{1k} (L^\dagger a_{1k}+a^\dagger_{1k}L)+\sum_k g_{2k}(L a_{2k}+a^\dagger_{2k}L^\dagger),
\label{h3}
\end{eqnarray}
where, $g_{1k}=g_k \cosh(\theta_{k})$ and $g_{2k}=g_k \sinh(\theta_{k})$.
The thermal vacuum 
can be written in terms of the vacuum for $b_k$, $c_k$ modes, $|\Omega_0\rangle$, as
\bea
|\Omega\rangle=e^{-iG}|\Omega_0\rangle.
\label{thvac}
\eea
The thermal vacuum can be written in alternative ways that further enlighten its physical meaning. Firstly, it can be written as $|\Omega\rangle=e^{-S/2}e^{\sum_k b_k^\dagger c_k^\dagger}|\Omega_0\rangle$, 
with $S=-\sum_k(b_k^\dagger b_k \log\sinh^2(\theta_{k})-b_kb_k^\dagger \log\cosh^2(\theta_{k}))$, which can be interpreted as the entropy operator for the physical (original) environment \cite{blasonebook}, 
since the thermofield vacuum is the state that minimizes the thermodynamic potential $\langle \Omega|(-\frac{1}{\beta}S+H)|\Omega\rangle$.
Secondly, up to normalization, $|\Omega\rangle\propto e^{-\beta H_B/2}|I\rangle$, where $|I\rangle=\sum_n|n\rangle_b |n\rangle_c$ is a maximally entangled state between the real and the auxiliary environments, defined in terms of their energy eigenstates, $|n\rangle_b $, $|n\rangle_c$. 
The thermal state of the original environment is thus  $\rho_B=\Ttr_{\aaux}[|\Omega\rangle\langle\Omega|]$ 
and it can be approximated by a MPO by evolving the maximally entangled state in imaginary time~\cite{verstraete2004,zwolak2004}. 
In contrast, the present approach is based on directly calculating the dynamics of the whole system
under the Hamiltonian (\ref{h3}), using the thermofield vacuum as (pure) initial state for both reservoirs.
Although this state is annihilated by $a_{1k}$ and $a_{2k}$, 
the number of physical particles has non-vanishing expectation value 
$n_k=\langle \Omega|b_k^\dagger b_k|\Omega\rangle=\sinh^2(\theta_{k})$.
Hence, solving the dynamics of the initial problem (\ref{h1}) with an initial condition $\rho_0^\ttot=\rho^S_0\otimes\rho^\tth_B$, with $\rho_0^S$ the initial state of the system, is equivalent to solving the dynamics with (\ref{h3}), but considering $\rho_0^\ttot=\rho^S_0\otimes|\Omega\rangle\langle \Omega|$ (see Fig. (\ref{thermofig}a) and (\ref{thermofig}b) respectively). 
We have described in detail the thermofield transformation for bosonic environments, but a similar Bogoliubov transformation can be proposed for fermionic reservoirs.
In that case \cite{blasonebook} we have $a_{1k}=e^{-iG}b_ke^{iG}=\cos(\theta_{k})b_k-\sin(\theta_{k})c^\dagger_k$, and $a_{2k}=e^{-iG}c_ke^{iG}=\cos(\theta_{k})c_k+\sin(\theta_{k}) b^\dagger_k$.
With this transformation, the Hamiltonian (\ref{h2}) is transformed into (\ref{h3}).

\textit{Chain representation--}
The Hamiltonian (\ref{h3}) represents an OQS interacting with two independent environments, having operators $a_{1k}$ and $a_{2k}$ respectively. The whole problem can be mapped into 
a one dimensional structure with the schematic form in Fig. (\ref{thermofig}c). 
In general, the environment oscillators in (\ref{h1}) form a quasi-continuum, so that the Hamiltonian can also be written as $H=H_S+\int_{0}^{1}dkg(k)(b(k)L^\dagger+L b(k)^{\dagger})+\int_{0}^{1}\omega(k)b(k)^{\dagger}b(k)$.
When the environment is in a Gaussian state, $\omega(k)$ and $g(k)$ enter the description of the OQS only through the spectral density, $J(\omega)$.
In this situation, one can always choose $\omega(k)=\omega_0 k$ (with $\omega_0$ an arbitrary constant that may be taken as one), and $\hat{g}(k)=\sqrt{J(\omega(k))}$.
Similarly, the continuum representation of (\ref{h3}) reads  
\begin{align}
\tilde{H}_{\ttot}&=H_S +\int_0^1 dk k \large(a_{1k}^\dagger a_{1k}-a^\dagger_{2k}a_{2k}\large)\cr
&+\int_0^1 dk  \large[\hat{g}_{1k} (L^\dagger a_{1k}+a^\dagger_{1k}L)+\hat{g}_{2k}(L a_{2k}+a^\dagger_{2k}L^\dagger)\large] \nonumber
\label{h4}
\end{align}
Thus, the spectral densities are $ J_1(k)=\hat{g}_1^2(k)=\sum_k (1+n(\omega(k)))J(\omega(k))$, and $ J_2(k)=\hat{g}_2^2(k)=n(\omega(k))J(\omega(k))$.
Then, using the unitary transformation discussed in \cite{prior2010,chin2010b}, new bosonic operators $B_n$ and $C_n$ can be defined for each reservoir, such that
\bea
a_{1k}=\sum_n U_{1n}(k)B_n, \quad
a_{2k}=\sum_n U_{2n}(k)C_n,
\label{eq:BCn}
\eea
where $U_{jn}(k)=g_j(k)\pi_{jn}(k)/\rho_{nj}$ ($j=1,2$). Here, $\pi_{jn}(k)$ are monic orthogonal polynomials that obey $\int_{0}^{1} dk J_j(k)\pi_{j,n}(k)\pi_{j,m}(k)=\rho_{nj}^2\delta_{nm}$, with $\rho^2_{nj}=\int_{0}^{1} dk J_j(k)\pi^2_{j,n}(k)$ \cite{chin2010b,chinbook2011}. 
Hence, the proposed transformation is also orthogonal, $\int dk U^*_{jn}U_{jm}=\delta_{nm}$. The transformed Hamiltonian can be written as
$H_{tot}=H_S +H_B+H_\iint$,
with 
$H_\iint=g_1(L^\dagger B_0+B_0^\dagger L)+g_2(L C_0+C_0^\dagger  L^\dagger)$, with $g_j=\rho_{j0}$,
and
\begin{align}
H_B=&\sum_{n=0,\cdots,M}(\alpha_{1,n} B_n^\dagger B_n-\alpha_{2,n}C_n^\dagger C_n
\nonumber \\
+&\sqrt{\beta_{1,n+1}}B^\dagger_{n+1}B_n-\sqrt{\beta_{2,n+1}}C^\dagger_{n+1}C_n+h.c.),
\end{align}
where the recurrence relation of the polynomials has been used,
$\pi_{j,n+1}(k)=(k-\alpha_{j,n})\pi_n(k)-\beta_{j,n}\pi_{j,n-1}(k)$, with $\pi_{j,n-1}=0$. 
Coefficients $\alpha_{j,n}$ and $\beta_{j,n}$ can be obtained with standard numerical routines \cite{gautschi2005}. 
The resulting Hamiltonian describes two tight-binding chains to which the system is coupled. The thermofield vacuum 
is also annihilated by the new modes $B_n$ and $C_n$, so that the dynamics of the whole system can be simulated 
using MPS time-evolution methods from an initial state with zero occupancy of each of these modes.

A similar mapping can be applied in the case of 
a finite discrete environment by means of 
a standard
tri-diagonalization.

In the following, we present numerical results to illustrate this approach in different examples.

\textit{Example 1: A spin in a bosonic field--}
Let us consider a spin $1/2$ system coupled to a bosonic environment with spectral density given by the 
Caldeira and Leggett model \cite{leggett1987,weissbook},
\begin{eqnarray}
J(\omega)=\eta \omega^s e^{-\omega/\omega_{c}},
\label{chapuno41}
\end{eqnarray}
with $0<s<1$ in the \textit{sub-ohmic} case, and $s>1$ in the \textit{super-ohmic}. Roughly speaking, the constant $\eta$ 
gives the coupling strength between system and environment. 
The exponential factor in (\ref{chapuno41}) provides a smooth cut-off for the spectral density, modulated by a frequency cut-off $\omega_{c}$. 
This general model provides a good approximation for spectral densities appearing in many different problems, like an impurity in a photonic crystal \cite{florescu2001,devega2005}, quantum impurity models \cite{si2001}, and solid state devices at low temperatures such as superconducting qubits \cite{shnirman2002}, quantum dots \cite{tong2006}, and nanomechanical oscillators \cite{seoanez2007}, to name just a few examples.
As a first check we consider a solvable example, with
$H_S=\frac{1}{2}\hbar \omega_S \sigma_z$, and $L=\sigma_z$ in (\ref{h1}). For an initial state $|\psi_0 \rangle=a| 0\rangle+b| 1\rangle$, the expectation value of any system operator, $A$, can be analytically calculated \cite{alonso2007},
\beqa
\langle A (t)\rangle&=&e^{-2 \phi_t}
\big\{
|a|^2 
e^{i \omega_St} 
+
|b|^2
e^{-i \omega_S t}
\big\},
\label{model5}
\eeqa
with $\phi_t=\int_0^t d\tau \int_0^\tau ds \Rre[\alpha_T(\tau-s)]$, $\alpha_T(t)=\sum_k g_k^2 \left[\coth{(\frac{\omega_k \beta}{2})}\cos{(\omega_k t)-i\sin{(\omega_k t)}}\right]$.
\begin{figure}[ht]
\centerline{\includegraphics[width=0.45\textwidth]{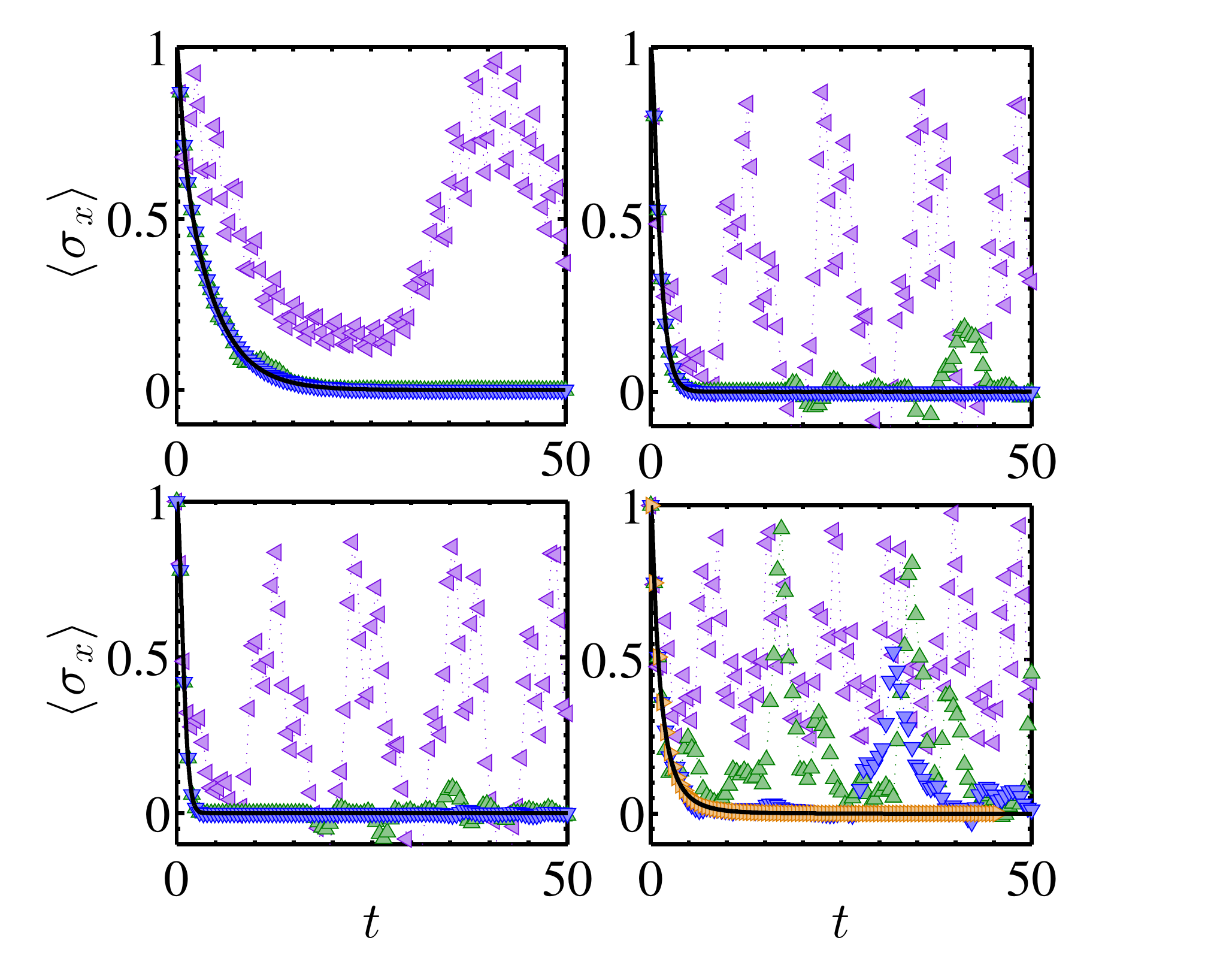}}
\caption{Evolution of the mean value of $\langle\sigma_x(t)\rangle$ for $a=b=1/\sqrt{2}$, for $\eta=0.1$ and $\omega_S=0$. Upper panels correspond to the ohmic model ($s=1$), for $\beta=5$ (left) and $\beta=1$ (right panel). The two lower panels correspond respectively to a sub-ohmic model with $s=1/2$ (left panel), and a super-ohmic model with $s=3/2$ (right panel), both for $\beta=1$. The exact solution is given by the solid black curves. For the MPS,  
we consider a varying $M$: magenta, green and blue curves correspond to $M=2,10,40$ respectively in all plots. The last plot includes an orange curve with $M=120$. Bond dimension is $D=20$, and maximum occupation numbers in the harmonic oscillator basis is $n=3$. \label{thermofig2}}
\end{figure}
To simulate the problem numerically using MPS we need to truncate 
the maximum occupation number of the bosonic modes, and 
the length of the chains corresponding to the transformed environment. We compare the numerical solution to the exact one for $A=\sigma_x$ in  Fig. \ref{thermofig2}. and observe very good agreement for all considered spectral densities, couplings and temperatures,  for a relatively small bond dimension and length of each chain, $M$.

In the following, we consider a problem that is not exactly solvable, by choosing $L=\sigma_x$, and compare the solutions of our method with those corresponding to a master equation (ME) up to second order in the system-environment coupling parameter $g$
\begin{eqnarray}
\frac{d\rho_s  (t)}{dt}&=&-i[H_S ,\rho_s (t) ]+\int_0^t d\tau \alpha_2^{*}(t-\tau) [L^\dagger ,\rho_s (t)  L(\tau-t)]\nonumber\\
&+&\int_0^t d\tau \alpha_2(t-\tau) [L^\dagger(\tau-t) \rho_s (t) ,L]\nonumber\\
&+&\int_0^t d\tau\alpha_1(t-\tau)[L(\tau-t)\rho_s (t) ,L^{\dagger}]\nonumber\\
&+&\int_0^t d\tau \alpha_1^{*} (t-\tau)[L,\rho_s (t) L(\tau-t)^{\dagger}]+{\mathcal O}(g^3),
\label{icc20chap3a}
\end{eqnarray}
with $\alpha_1 (t-\tau)=\sum_k g^2_k (n_k+1)e^{-i\omega_k (t-\tau)}$, $\alpha_2 (t-\tau)=\sum_\lambda g^2_k n_ke^{i\omega_k (t-\tau)}$, and $L(t)=e^{iH_S t}Le^{-iH_S t}$. 
\begin{figure}[ht]
\centerline{ \includegraphics[width=0.3\textwidth]{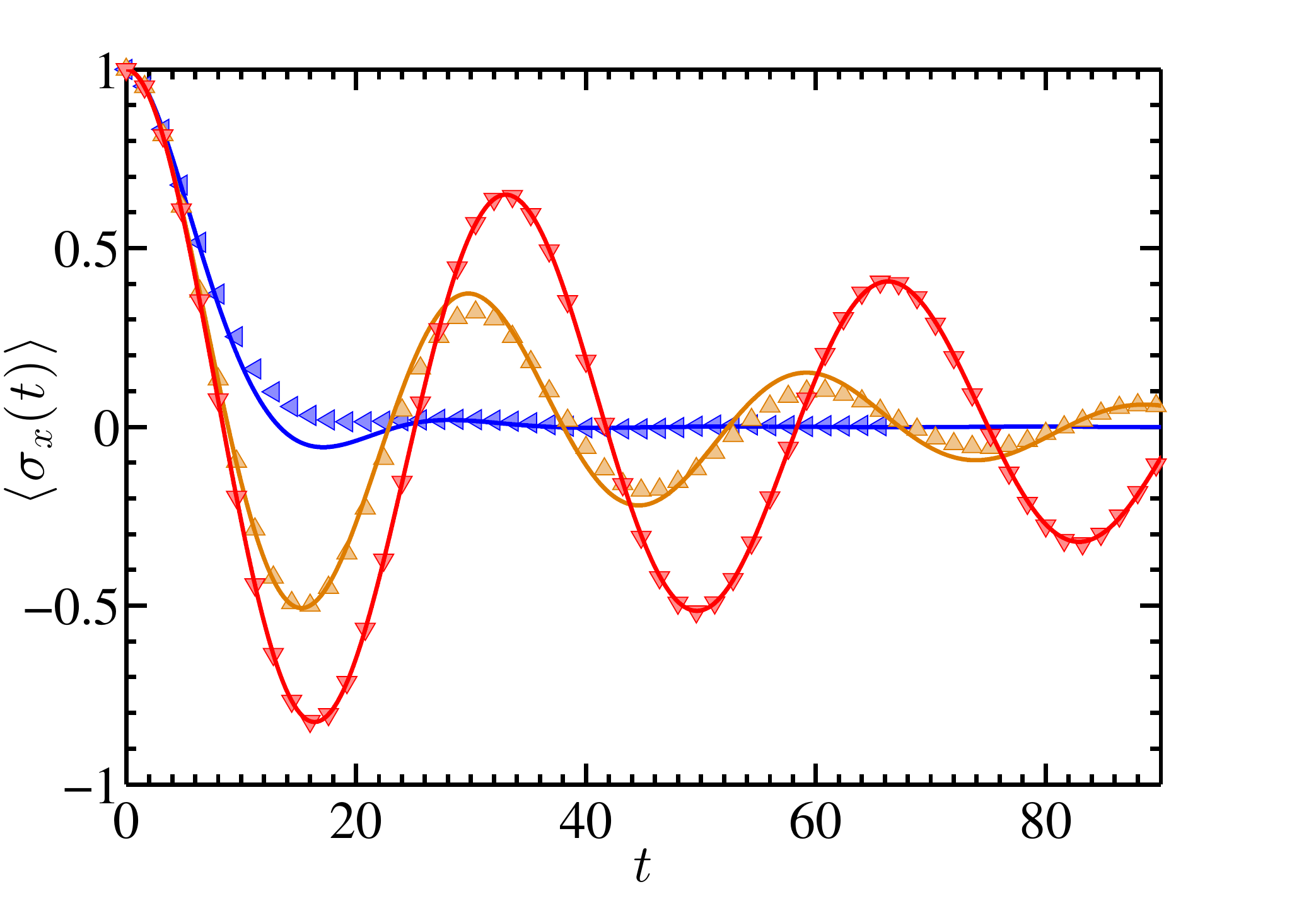}}
\caption{Comparison of ME (solid curves) with MPS (triangles) results for $\langle\sigma_x(t)\rangle$ considering $\eta=0.01$, $\omega_S=0.1$, and $M=100$ for both chains. We consider $\beta=10$ (blue), $\beta=50$ (orange) and $\beta=1000$ (red). The MPS results converge with maximum population per oscillator $n=4$. \label{thermofig3}}
\end{figure}
To derive this equation, the Born approximation has also been assumed. This ME neglects the system-environment correlations, and considers that the latter remains in the thermal equilibrium state $\rho_B$ during the interaction, so that $\rho_\ttot(t)\approx\rho_s(t)\otimes\rho_B$. 

\begin{figure}[ht]
\centerline{ \includegraphics[width=0.32\textwidth]{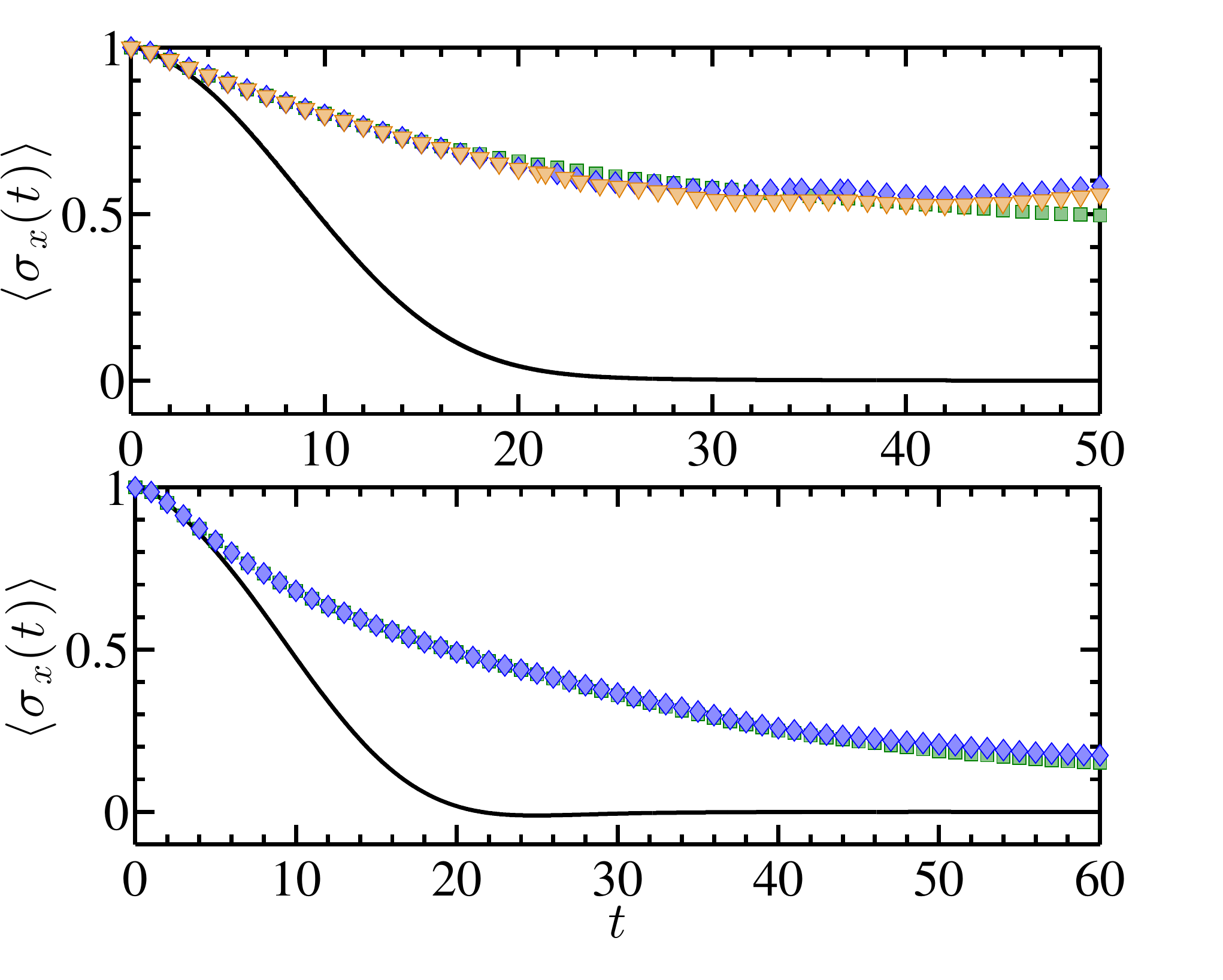}}
\caption{Evolution of $\langle\sigma_x(t)\rangle$ for $\beta=10$ (upper panel, with maximum bond dimension $D=40$) and $\beta=50$ (lower panel, with $D=20$) for $\eta=0.1$, $\omega_S=0.1$, and $M=100$ for both chains. 
The solid black curve corresponds to the solution for the ME. Considering $n_1$ the dimension of the first two oscillators in the chain, and $n_2$ the dimension of the following ones, the curves with green squares 
correspond to $(n_1=5,n_2=4)$, and the ones with blue diamonds to $(n_1=6,n_2=5)$ (lower panel) and $(n_1=7,n_2=6)$ (upper panel). The curve with orange triangles in the upper panel corresponds to $(n_1=8,n_2=7)$. 
\label{thermofig4}}
\end{figure}

As shown in Fig. (\ref{thermofig3}), the ME and the MPS coincide quite reasonably at weak couplings. However, as shown in Fig. (\ref{thermofig4}), for stronger couplings the ME does not give an accurate description of the dynamics.  Indeed, the MPS results describe comparatively a much slower decay for the two temperature values here considered.
Also, the computational cost of the MPS in the strong coupling regime is much higher than at weak coupling. 
Nevertheless, the difference of the present scheme is that \textit{the excitations involved in the numerical resolution are just those that are dynamically generated due to the interaction with the OQS}. This is in clear contrast with traditional methods in which the initial state is thermal, and therefore already has a finite initial occupation in the environment basis. 

\textit{Example 2: A quantum dot coupled to an electronic reservoir--} 
As noted above, our proposal is valid also for fermionic environments. To illustrate this, we consider a quantum dot (QD) coupled to an electronic reservoir at a finite temperature with a Hamiltonian $H = H_{S} + H_{\rm B} + H_{\rm hy}$. Here, 
$H_{S} = \sum_{\sigma} (V n_{\sigma} +  \frac{U}{2} ~n_{\sigma} n_{\bar\sigma } )$ is the Hamiltonian of the quantum dot, which is represented using the Anderson impurity model with an on-site Coulomb repulsion $U$ and an on-site energy $V$. Here, the operator $n_{\sigma} = d^\dagger_{\sigma} d_{\sigma}$ measures the number of electrons with spin $\sigma=\uparrow,\downarrow$ at the dot.
We consider that the QD is connected to the reservoir through a hybridization term $H_{\rm hy} =- t\sum_{k;\sigma} g_k(d^\dagger_{\sigma} b_{k} + {\rm h.c.} )$, that is a sum of bilinear terms wherein $d^\dagger_{i\sigma}$($d_{i\sigma}$) creates (annihilates) an electron at the dot with spin $\sigma$ and $b^\dagger_{k}$($b_k$) creates (annihilates) an electron with arbitrary spin and momentum $k$ in the reservoir.  Hence, the interaction Hamiltonian has a similar form as the one in (\ref{h1}), but redefining $L=-t\sum_\sigma d_\sigma$. For simplicity, we have considered that both spins $\sigma$ couple equally to the reservoir. The Hamiltonian of the environment is  $H_{\rm B} =\sum_k \omega_k b_{k}^\dagger b_{k}\label{Hleads}$.
After the thermofield transformation, the former Hamiltonian is written in terms of 
$\tilde{H}_{\rm B}=\sum_k\omega_k (a_{1k}^\dagger a_{1k}-a_{2k}^\dagger a_{2k})$, 
and an interaction Hamiltonian of the form (\ref{h3}) with couplings
$g_{1k}=- tg_k\sqrt{1+f_k}$, and $g_{2k}=- tg_k\sqrt{f_k}$, with $f_k=(1+\exp(\beta\omega_k))^{-1}$. We consider a spectral density of sub-ohmic type, with $s=0.5$ in Eq. (\ref{chapuno41}).

\begin{figure}[ht]
\centerline{ \includegraphics[width=0.32\textwidth]{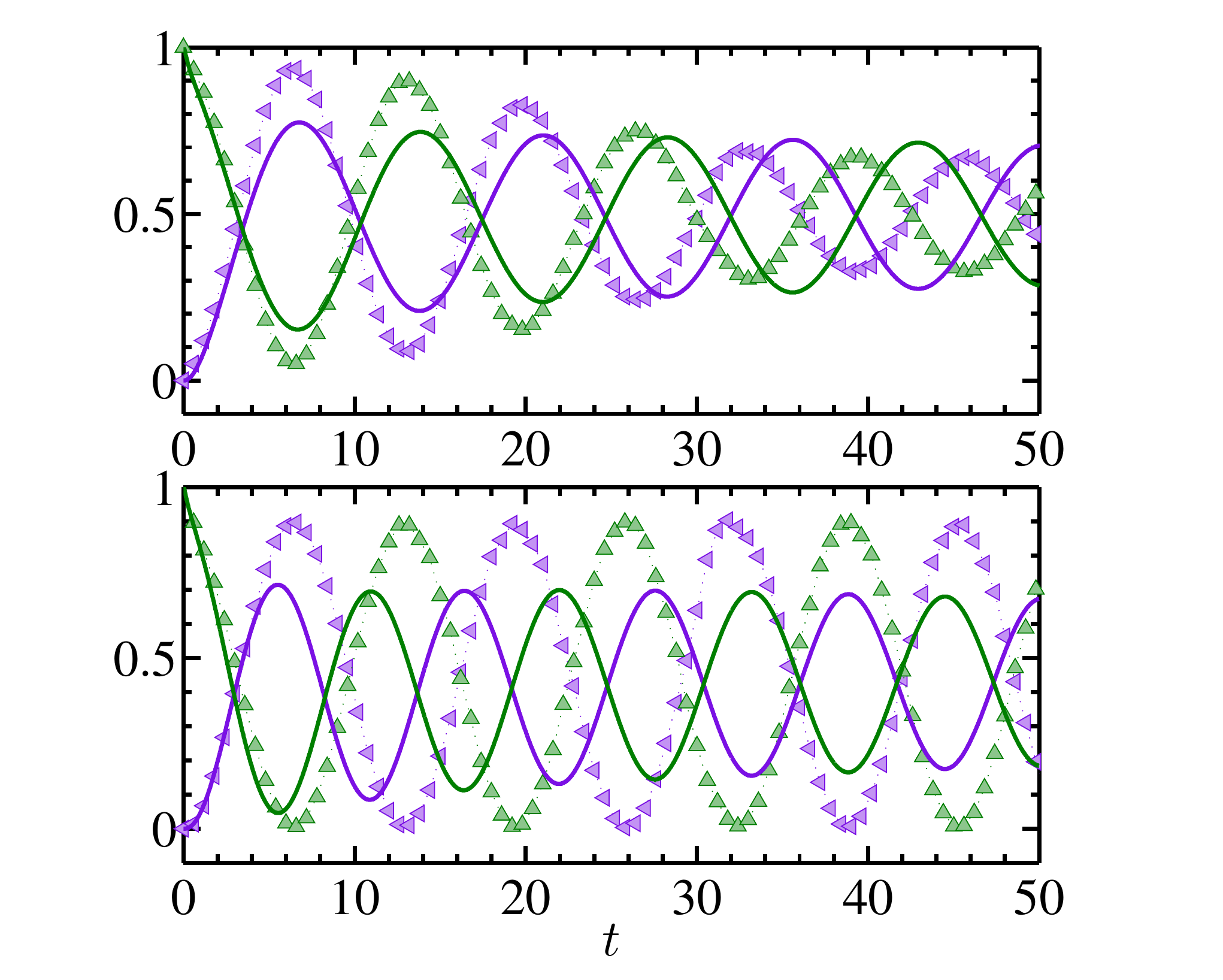}}
\caption{Evolution of $\langle n_\uparrow\rangle$ (green) and $\langle n_\downarrow\rangle$ (purple) for the ME (solid lines) and the t-DMRG (symbols). We have considered $\beta=1$ (upper panel), $\beta=10$ (lower panel) with $U=0.2$, $V=-U/2$, $\omega_c=15$, $t=0.01$, and a $M=100$ oscillators in the chain.
 \label{thermofig5}}
\end{figure}

Comparing the MPS results to those of ME, 
as shown in Fig. \ref{thermofig5},
we find initial agreement as expected, but then the results start to differ considerably even at relatively weak couplings. 
Due to the limited size of the fermionic basis, the MPS converges to the exact result with relatively small computational resources. 
\textit{Conclusions and outlook--} Based on a thermofield approach, our formalism allows us to efficiently integrate the dynamics of an OQS coupled to a thermal reservoir, either bosonic or fermionic, in a pure state formalism,
without previously preparing the thermal state with imaginary time evolution. The approach is based on performing an analytical (thermal Bogoliubov) transformation over the (physical) environment and an auxiliary one. 
Provided the thermal state of the original environment is known, more concretely, that the quantities $n_k$ can be analytically or numerically computed, 
our approach can be used to solve thermalization problems of OQS using only zero-temperature (pure state) MPS. 

{\it Acknowledgment:} 
We thank D. Alonso, U. Schollw\"ock, F. Heidrich-Meisner, and C.A. B\"usser for helpful discussions. 
This work was supported by Nanosystems Initiative Munich (NIM) (project No. 862050-2), and partially by the Spanish MICINN (Grant No. FIS2013-41352-P), and the EU through SIQS grant (FP7 600645).
\bibliography{/Users/ines.vega/Dropbox/Bibtexelesdrop}
\bibliographystyle{prsty}

\end{document}